\documentclass[twocolumn,showpacs,preprintnumbers,amsmath,amssymb]{revtex4-1}

\usepackage{graphicx}
\usepackage{dcolumn}
\usepackage{bm}
\usepackage{rotating}
\usepackage{amssymb}

\begin{document}


\title{Triplet-singlet conversion in ultracold Cs$_2$ and production of ground state molecules}

\author{Nadia Bouloufa$^{1}$}
\author{Marin Pichler$^{2}$}
\author{Mireille Aymar$^{1}$}
\author{Olivier Dulieu$^{1}$}
\affiliation{$^{1}$Laboratoire Aim\'e Cotton, CNRS, Universit\'e Paris-Sud, B\^at. 505, 91405 Orsay, France}
\affiliation{$^{2}$Physics Department, Goucher College, Baltimore MD, 21204}

\date{\today}

\begin{abstract}
We propose a process to convert ultracold metastable Cs$_2$ molecules in their lowest triplet state into (singlet) ground state molecules in their lowest vibrational levels. Molecules are first pumped into an excited triplet state, and the triplet-singlet conversion is facilitated by a two-step spontaneous decay through the coupled $A^{1}\Sigma_{u}^{+} \sim b ^{3}\Pi_{u}$ states. Using spectroscopic data and accurate quantum chemistry calculations for Cs$_2$ potential curves and transition dipole moments, we show that this process has a high rate and competes favorably with the single-photon decay back to the lowest triplet state. In addition, we demonstrate that this conversion process represents a loss channel for vibrational cooling of metastable triplet molecules, preventing an efficient optical pumping cycle down to low vibrational levels.
\end{abstract}


\maketitle

\section{Introduction}
\par
Research on cold and ultracold molecules  currently attracts considerable interest due to their features and applications in fundamental science \cite{Carr2009}, from the control of low-energy quantum dynamics of few-body systems, to the intrinsic many-body nature of quantum degenerate gases \cite{Goral2002}, as well as due to the prospects in applied areas such as quantum computing \cite{DeMille2002}, precision measurements \cite{DeMille2008, Hudson2006}, and ultracold controlled chemistry \cite{FaradayDisc2009}. Therefore work on creating samples of cold molecules has rapidly progressed, and recent reviews  outline the experimental methods, theoretical issues and applications for neutral \cite{Carr2009,ODulieu2009, Kremsbook2009} and charged species \cite{Willitsch2008,Gerlich2008a,Gerlich2008b}. Except in remarkable cases \cite{Stuhl2008, Shuman2009}, the rich internal structure of molecules prevent them to be directly cooled down by lasers.

Among the main challenges is the formation of ultracold molecules in a well-defined internal quantum state, which would open the way for the full control of its evolution under various interactions. Therefore work on creating samples of ultracold molecules has rapidly progressed in two main directions: either molecules are created by association of ultracold atom pairs induced by a photon (photoassociation, or PA) \cite{Jones2006} or by a magnetic field (magnetoassociation, or MA) \cite{Kohler2006,Chin2010}, or by direct cooling applied to pre-existing molecules, such as buffer gas cooling \cite{Doyle2004} or Stark deceleration \cite{Meerakker2008}. The former approach yields molecular samples with temperatures in the micro- or nanokelvin range, but most often with high internal vibrational energy. In contrast the latter approach produces molecules in their lowest energy levels, but their translational motion is characterized by a temperature larger than 1 millikelvin in most cases. The ultimate goal to produce ultracold molecules in micro- or nano-Kelvin range with no vibrational or rotational excitation, or in a well-defined internal quantum state still poses a difficult challenge. Using stimulated Raman adiabatic passage (STIRAP) technique, the population of high-lying bound levels of molecules created by MA of ultracold atoms in a degenerate or near-degenerate quantum gas has been transferred down to the lowest molecular bound level through a single STIRAP step in KRb \cite{Ni2008} and through a double STIRAP step in Cs$_2$ \cite{Danzl2008,Danzl2010}. These results pave the path towards the achievement of a degenerate gas of molecules with no internal energy.  Additionally, ultracold alkali-metal diatomic molecules produced by PA have been transferred into their lowest vibrational level $v=0$ by stimulated emission pumping for RbCs \cite{Sage2005} and LiCs \cite{Deiglmayr2008} and by optical pumping for Cs$_2$ \cite{Viteau2008}. However the relatively small fraction (a few thousands) of ground state molecules in the ultracold gas could be a limitation for further studies.

\begin{figure}
    \begin{center}
    \includegraphics[width=1\linewidth]{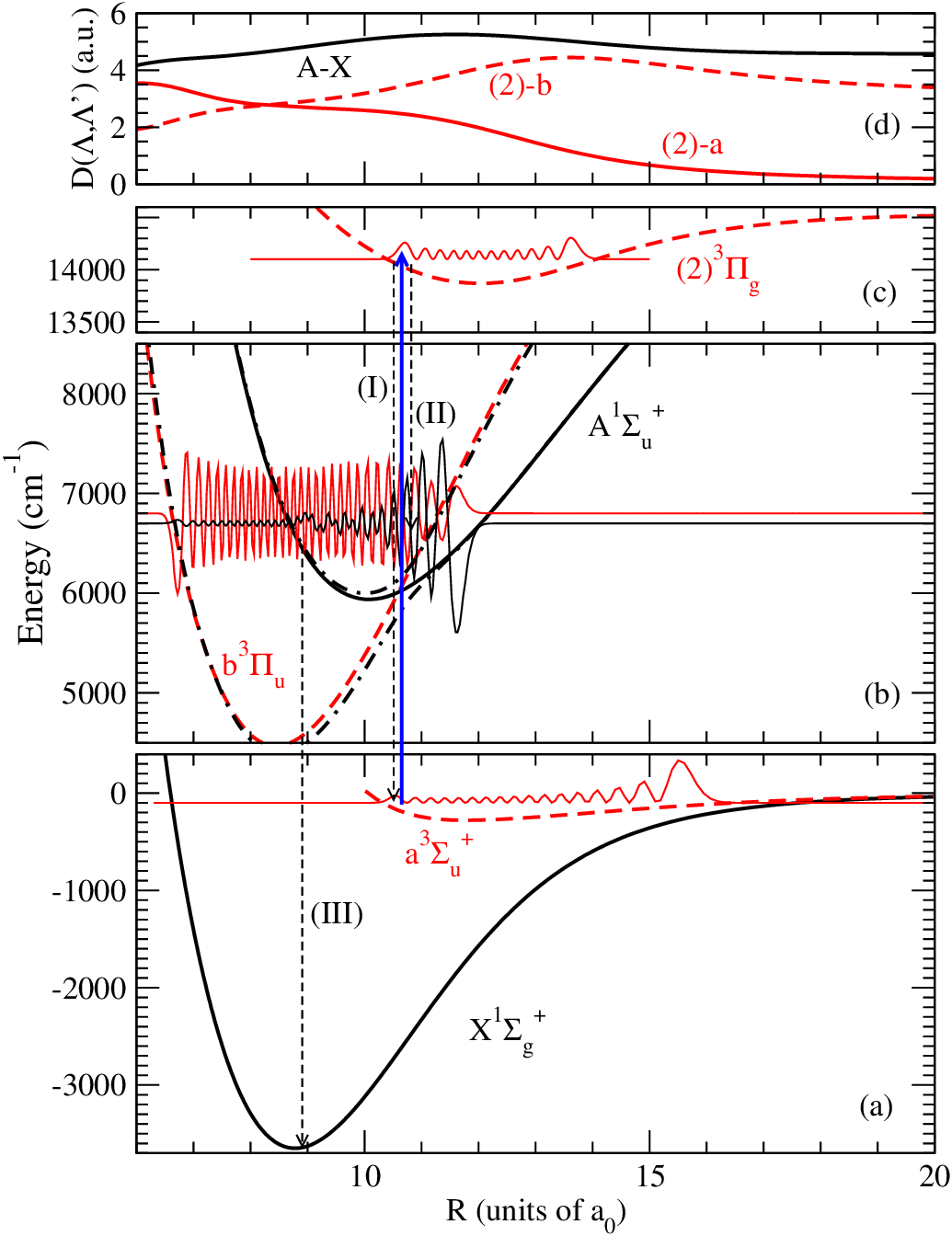}
    \end{center}
    \caption{(Color on line) Scheme of the proposed process. (a) the $a^{3}\Sigma_{u}^{+}$ $v=17$ level is initially populated by PA. (c) It can be efficiently excited (thick vertical arrow) to the $(2)^{3}\Pi_{g}$ $v=12$ level. (b) the $A^{1}\Sigma_{u}^{+}(6s+6p)$ and $b^{3}\Pi_{u}(6s+6p)$ potential curves, and the resulting $0_u^+$ curves (dot-dashed lines) after diagonalization of the potential matrix of the $A$ and $b$ curves coupled by the spin-orbit coupling function of ref.\cite{SpiesPhD}. The $(2)^{3}\Pi_{g}$ level can decay either directly to the $a$ state (arrow I), or to the $X$ state via the $0_u^+$ states (arrow II and III). Probability densities are reported with an arbitrary scale. For the $0_u^+$ coupled wave function, the component on the $b$ state (upper red trace) and on the $A$ state (lower black trace) of the wave function are both represented on the same scale. (d) $R$-dependent dipole moment for the transitions $a^{3}\Sigma_{u}^{+}$-$(2)^{3}\Pi_{g}$ ($D_{(2)-a}$), $b^{3}\Pi_{u}$-$(2)^{3}\Pi_{g}$ ($D_{(2)-b}$), and $X^{1}\Sigma_{g}^{+}$-$A^{1}\Sigma_{u}^{+}$ ($D_{A-X}$). }
    \label{fig:transfer_scheme}
\end{figure}

In all the cases above, coupling between excited electronic state mediating the population transfer plays a crucial role, as it has been previously demonstrated in several PA experiments \cite{Dion2001,MPichler2006A,Bergeman2006,Haimberger2009}, and some other possibilities are still to be discovered \cite{Stwalley2010}. Typically, samples of ultracold molecules are initially created in their lowest metastable triplet electronic state, and are converted into ground state molecules via a transfer mechanism relying on a triplet-singlet coupling in excited electronic states induced by spin-orbit interaction. In the present paper, we propose a triplet-singlet conversion mechanism in Cs$_2$ which takes advantage of the large formation rate of triplet molecules that are indeed produced by PA of ultracold cesium atoms \cite{Fioretti1998,Drag2000}. A one-photon transition excites molecules in the lowest $a^{3}\Sigma_{u}^{+}$ state into a well-defined level of the excited $2^{3}\Pi_{g} (6s+5d)$ state which relaxes down into low vibrational levels of the ground state in two steps via the  $0_{u}^{+}(A^{1}\Sigma_{u}^{+} \sim b ^{3}\Pi_{u})$ coupled states (Fig.\ref{fig:transfer_scheme}). We show that the probability for accumulating molecules in their ground X$^{1}\Sigma_{g}^{+}$ state is comparable to the one for accumulating of molecules back into the $a^{3}\Sigma_{u}^{+}$ state. Besides the triplet-singlet conversion of cold molecules, this mechanism allows for $u-g$ symmetry conversion, which cannot be achieved via a two-photon process in homonuclear molecules, in contrast with heteronuclear molecules. As a further result, this conversion process suggests a likely explanation for the suppression of vibrational cooling of $a^{3}\Sigma_{u}^{+}$ molecules down to low vibrational levels as reported in ref.\cite{Sofikitis2010}.

The proposed model is described in Section \ref{sec:model} and the transfer efficiency is evaluated in Section \ref{sec:twostep}. The competition with other possible decay channels is discussed in Section \ref{sec:discussion} where prospects for experimental realization of the proposed scheme are considered.

\section{The proposed triplet-singlet conversion scheme}
\label{sec:model}

We investigate the efficiency of the process depicted in Fig.\ref{fig:transfer_scheme}.  Our model is based on accurate spectroscopic information recently obtained on Cs$_2$ electronic states. The X$^{1}\Sigma_{g}^{+}$ ground state potential curve is taken from ref.\cite{Amiot2002}, while the lowest $a^{3}\Sigma_{u}^{+}$ potential curve comes from the recent analysis of Li Li's group \cite{Xie2009}. The triplet-singlet conversion is mediated by the $A^{1}\Sigma_{u}^{+}(6s+6p)$ and $b^{3}\Pi_{u}(6s+6p)$ states (hereafter referred to as the $A$ and $b$ states, respectively) coupled by a spin-orbit (SO) interaction, which depends on the internuclear distance $R$ \cite{Kokoouline2000}. According to Hund's case c labeling, these coupled states result in a pair of $0_u^+$ states dissociating into the $6^2S_{1/2}+6^2P_{1/2,3/2}$ limits. In the following we will refer to the $0_u^+(A \sim b)$ pair of states (see dotted lines in Fig.\ref{fig:transfer_scheme}). The potentials curves for the $A$ and $b$ states are calculated by \textit{ab initio} methods described in ref.\cite{Aymar2005}, which were adjusted empirically to reproduce the spectroscopic data of references \cite{Verges1987} and \cite{Xie2009}. The SO coupling function is taken from the \textit{ab initio} determination of ref.\cite{SpiesPhD}. The $(2)^{3}\Pi_{g}$  potential curve as well as the $R$-dependent transition dipole moments are also evaluated according to the method of ref.\cite{Aymar2005} (Fig.\ref{fig:transfer_scheme}d). The $(2)^{3}\Pi_{g}$  potential curve is found with an harmonic constant $\omega_e=17.1$~cm$^{-1}$ in good agreement with the measurement or ref.\cite{Diemer1991}; it is shifted downwards by 112~cm$^{-1}$ to match its  minimum with the value reported in ref.\cite{Diemer1991} for the $(2)^{3}\Pi_{g}(1_g)$ state. The vibrational energies and wave functions for all these molecular states are computed with the Mapped Fourier Grid Representation (MFGR) method \cite{Kokoouline1999}.

\begin{figure}
    \begin{center}
    \includegraphics[width=0.45\textwidth]{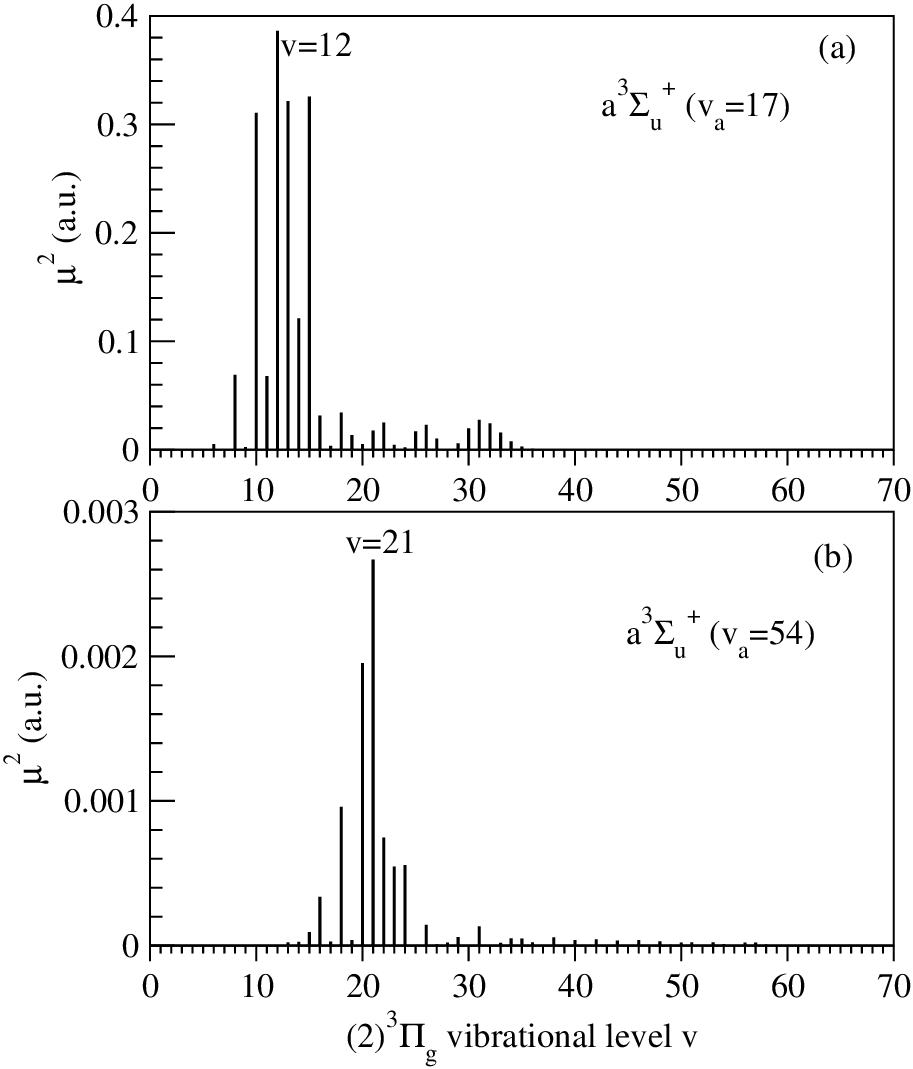}
    \end{center}
    \caption{Squared matrix element $\mu^2=| \langle a^{3}\Sigma_{u}^{+}, v_a | D_{(2)-a} | (2)^{3}\Pi_{g},v \rangle |^{2}$ (in atomic units) of the transition dipole moment function $D_{a-(2)}$ (see Fig.\ref{fig:transfer_scheme}d). (a): $v_a=17$; (b): $v_a=56$.}
    \label{fig:a_to_Pi}
\end{figure}

We assume that Cs$_2$ molecules are initially created by PA of cold Cs atoms into the so-called giant G1 or G3 resonances (at 11 720 and 11 715~cm$^{-1}$) characterized in ref.\cite{Vatasescu2006,Bouloufa2010} of the double-well $0_g^-(6^2S+6^2P_{3/2})$ state. These photoassociated molecules are stabilized by spontaneous emission in a distribution of vibrational levels of the $a^{3}\Sigma_{u}^{+}(6s+6s)$ (hereafter referred to as the $a$ state) located around the $v"=17$ level, as discussed in ref.\cite{Bouloufa2010}. These molecules are efficiently excited towards low-lying vibrational levels of the $2^{3}\Pi_{g}(6s+5d)$ state. Starting from one of the most populated ($v_a=17$) level of the $a$ state, the strongest transition is found towards the $v=12$ level of the $(2)^{3}\Pi_{g}$ state at 710~nm or 14084.6~cm$^{-1}$. This is illustrated in Fig. \ref{fig:a_to_Pi}a where the squared matrix element $| \langle a^{3}\Sigma_{u}^{+}, v_a=17 | D_{(2)-a} | (2)^{3}\Pi_{g},v \rangle |^{2}$ is shown. Then two electronic relaxations are open: either a direct transition back to the $a$ state, or a two-step transition down to the $X$ state via the $A-b$ coupled states. The arrows illustrating these transitions in Fig.\ref{fig:transfer_scheme} suggest that the classical turning points of the potentials match well enough to ensure that these two ways compete together, as confirmed in the next sections.

\section{The two-step spontaneous decay of the $2^{3}\Pi_{g}$ state}
\label{sec:twostep}

In this section, we demonstrate that the $v=12$ level of the $2^{3}\Pi_{g}$ state can efficiently create ground state molecules in their lowest vibrational level $v"=0$. We consider the problem only from the point of view of the transitions between vibrational levels, where transition strengths depend only on vibrational wave functions. We left apart the contribution of the rotational state of the molecules, which could be incorporated through H\"onl-London factors, without significantly modifying the main conclusions of the paper. The decay rate of a vibrational level $v_i$ of an electronic state $\Lambda$ towards all the vibrational levels $v_j$ of an electronic state $\Lambda'$ is given by the general expression for the Einstein coefficient $A_{v_i}(\Lambda-\Lambda')$ (in s$^{-1}$):
\begin{equation}
A_{v_i}(\Lambda-\Lambda')=\sum_{j}\frac{16\pi^3}{3\epsilon_0 c^3 h} \nu_{ij}^3 | \langle v^{\Lambda}_i |D_{\Lambda-\Lambda'}|v^{\Lambda'}_j \rangle|^2
\label{eq:rate}
\end{equation}
where $h\nu_{ij}$ is the energy difference between the $v_i$ and $v_j$ levels, and $D_{\Lambda-\Lambda'}(R)$ the $R$-dependent transition dipole moment between the electronic states $\Lambda$ and $\Lambda'$.

\begin{figure}
    \begin{center}
    \includegraphics[width=0.45\textwidth]{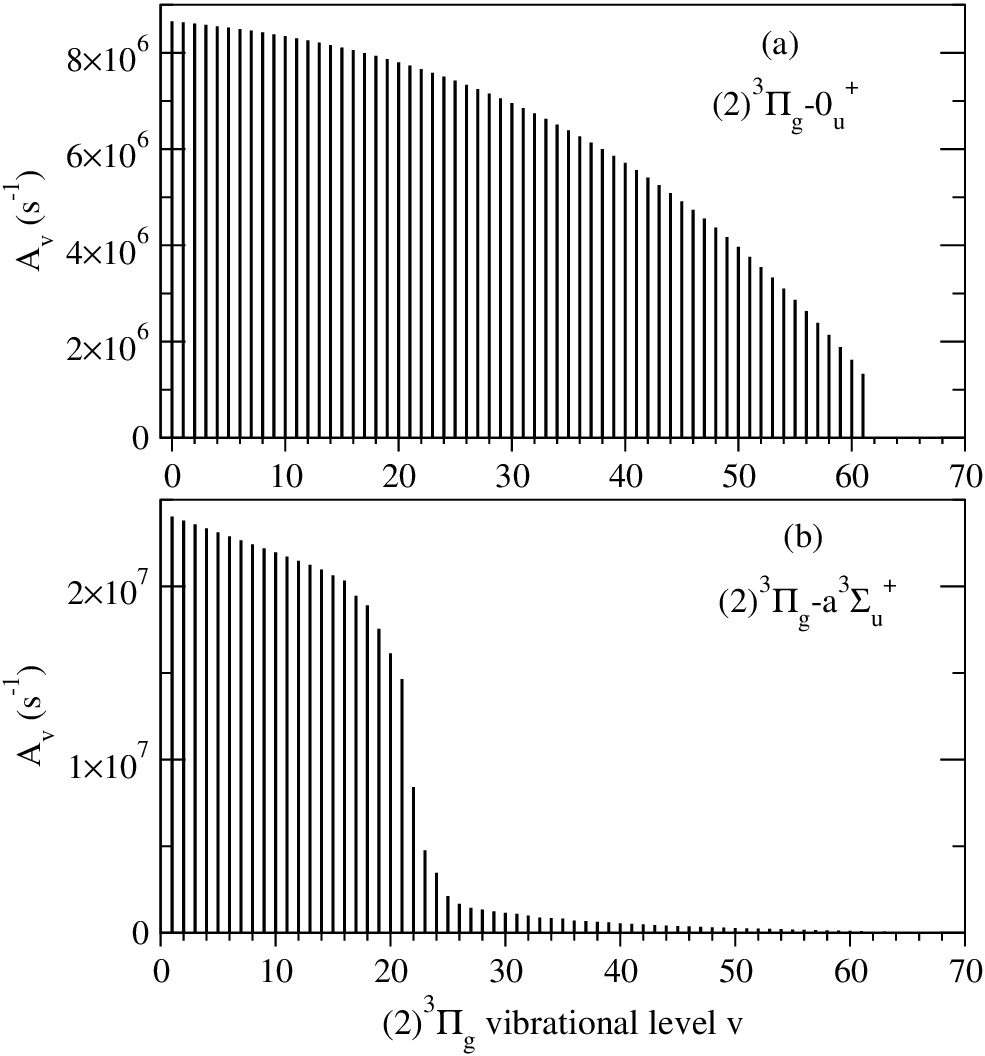}
    \end{center}
    \caption{Relaxation rate $A_v$ (in s$^{-1}$) of the $2^{3}\Pi_{g}$ vibrational levels (a) towards the levels of $0_{u}^{+} (A-b)$ system, and (b) towards the levels of the $a ^{3}\Sigma_{u}^{+}$ state.}
    \label{fig:2tPig_relax}
\end{figure}

\begin{figure}[t]
    \begin{center}
    \includegraphics[width=0.45\textwidth]{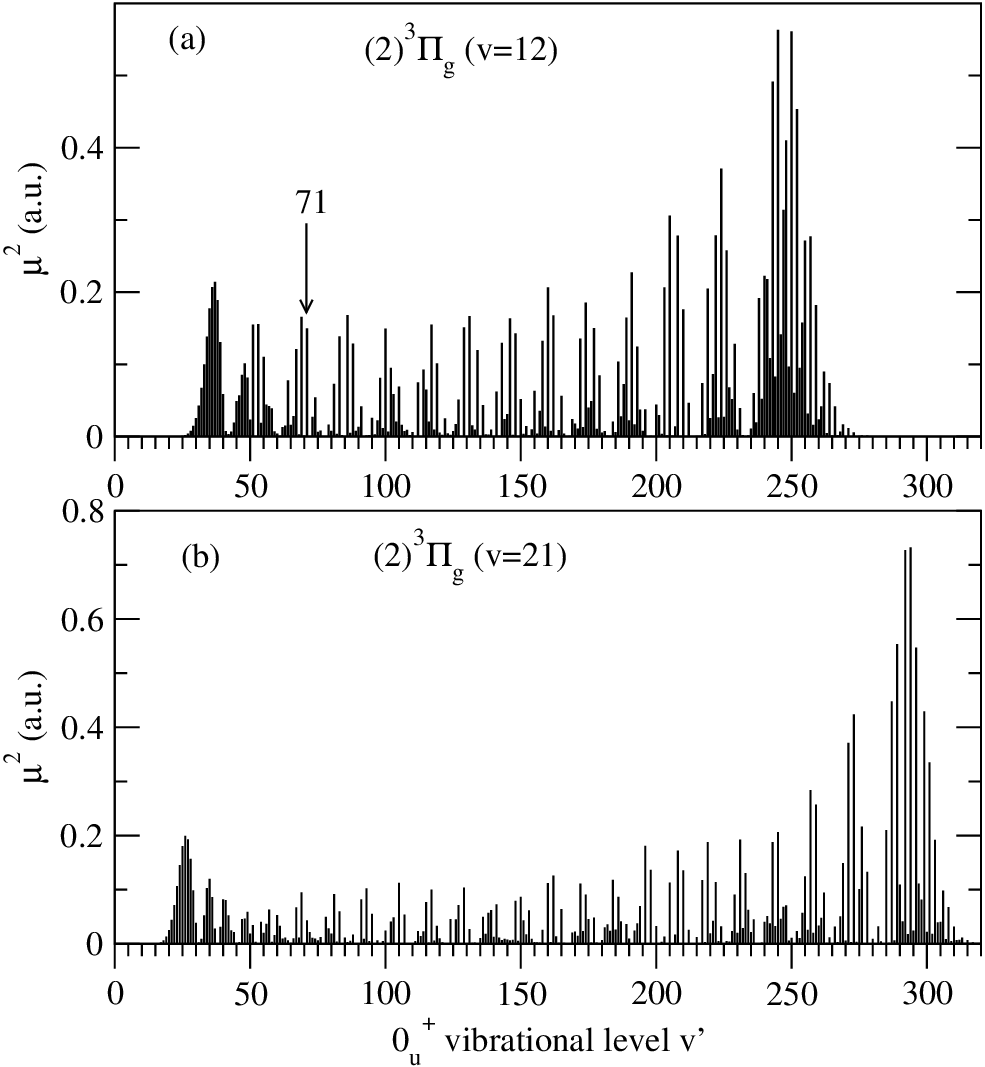}
    \end{center}
    \caption{Squared matrix element $\mu^2=| \langle 0_{u}^{+}, v' | D_{(2)-b} | (2)^{3}\Pi_{g}, v \rangle |^{2}$ (in atomic units) of the transition dipole moment function $D_{b-(2)}$ (see Fig.\ref{fig:transfer_scheme}d). The level $v'=71$ discussed in the text is indicated for clarity. (a) $v=12$; (b) $v=21$. }
    \label{fig:Piv6_to_0+system}
\end{figure}

We first compute the relaxation rate of a given $(2)^{3}\Pi_{g}$ vibrational level down to the $0_u^+(A\sim b)$ levels. This rate obviously depends on the amount of $b$ electronic character for each of the bound levels of the $0_u^+(A\sim b)$ coupled states, and on the dipole moment function $D_{b-(2)}$ (Fig.\ref{fig:transfer_scheme}d) for the transition $(2)^{3}\Pi_{g}$-$b^{3}\Pi_{u}$. We see in Fig. \ref{fig:2tPig_relax}a that $(2)^{3}\Pi_{g}$ $v=12$ level has a relaxation rate of $9\times10^{6} s^{-1}$ over the entire $0_{u}^{+}$ system. The population of the $0_{u}^{+}$ levels $v'$ after this relaxation is related to the squared matrix elements $| \langle 0_{u}^{+}, v' | D_{(2)-b} | (2)^{3}\Pi_{g}, v=12 \rangle |^{2}$ displayed in Fig.\ref{fig:Piv6_to_0+system}a. The envelope of this function reflects the oscillatory structure of the $v=12$ wave function. The $v'=71$ level of the coupled system corresponds to a local maximum in a range which is predicted below to decay favorably down to the $v"=0$ level of the $X$ state. This is also illustrated in Fig.\ref{fig:transfer_scheme} where it is shown that one of the turning points of the $v'=71$ wavefunction indeed coincides with the minimum of the $X$ potential curve. We note that the double STIRAP scheme of ref.\cite{Danzl2010} actually used the $0_u^+(v'=63)$-$X(v"=0)$ for their final step of the transfer.

\begin{figure}[t]
    \begin{center}
    \includegraphics[width=1\linewidth]{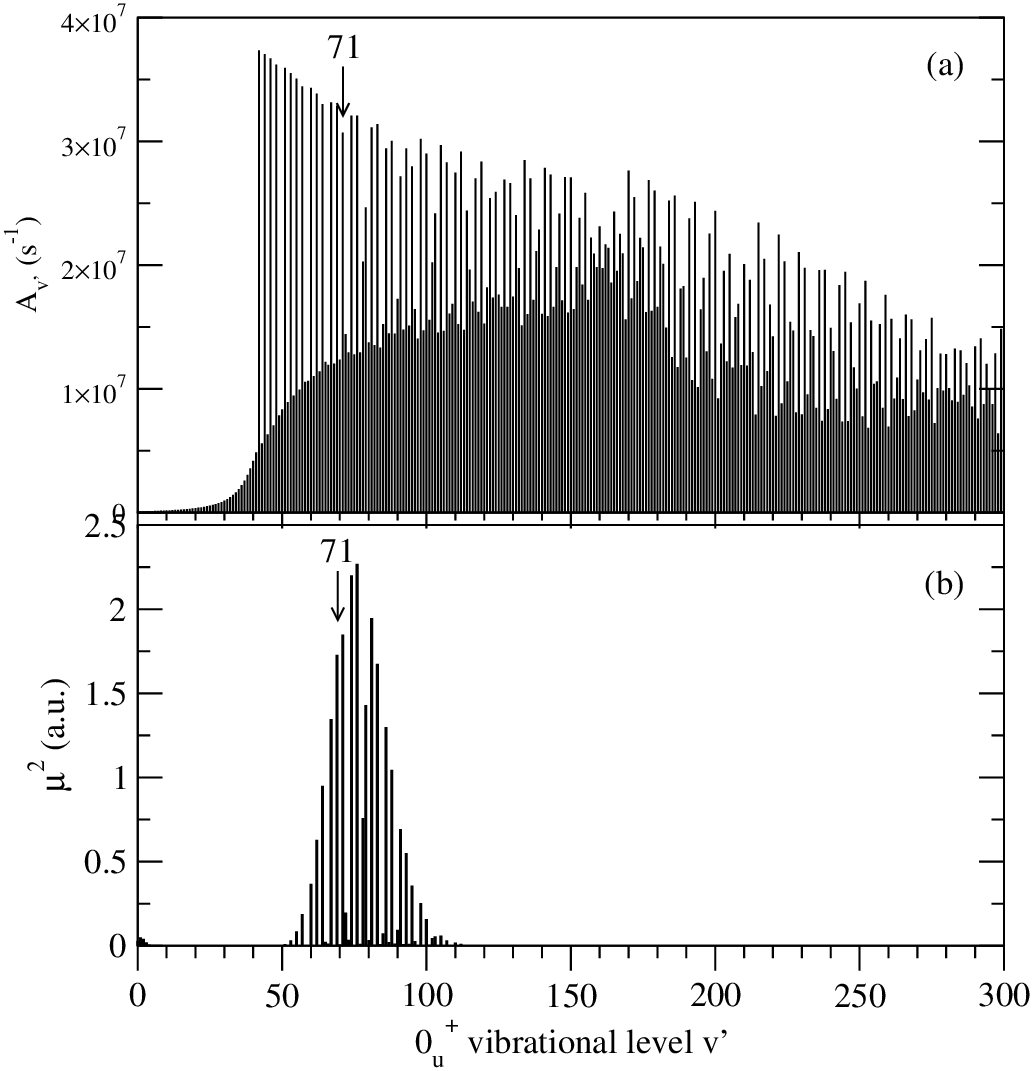}
    \end{center}
    \caption{(a) Relaxation rates $A_{v'}$ (in s$^{-1}$) of the vibrational levels of the $0_{u}^{+}$ coupled states towards the levels of the $X$ ground state. The $v'=71$ level is indicated for clarity. (b) The squared matrix element $| \langle X^{1}\Sigma_{g}^{+} v"=0 | D_{A-X} | 0_{u}^{+},v' \rangle |^{2}$, indicating that the $v'=71$ $0_{u}^{+}$ level indeed decay efficiently to $v"=0$.}
    \label{fig:0u+system_to_X}
\end{figure}

Using an expression similar to Eq.\ref{eq:rate}, we display in Fig. \ref{fig:0u+system_to_X}a the decay rate of the $0_{u}^{+}$ levels down to the ground state. It depends on the squared matrix element $| \langle X^{1}\Sigma_{g}^{+} v" | D_{A-X} | 0_{u}^{+},v' \rangle |^{2}$. Below $v'= 42$, \textit{i.e.} below the crossing between the $A$ and $b$ potential curves, $0_{u}^{+}$ levels have mostly a $b^{3}\Pi_{u}$ character, so that their decay rate towards $X$ levels is very low. Above this region vibrational levels acquire a significant $A^{1}\Sigma_{u}^{+}$ character whose amplitude oscillates from one level to another (see for instance ref.\cite{Kokoouline2000}). Such levels contribute to the dark background of lines in Fig. \ref{fig:0u+system_to_X}a. Among them, the levels which have the largest overlap with $X$ wave functions (like $v"=71$) are visible as lines with an amplitude about two times larger than the one of the background, and 4 to 5 times larger than that of the $(2)^{3}\Pi_{g}, v=12$-$0_u^+(A\sim b)$ decay rate. Figure \ref{fig:0u+system_to_X}b shows that the squared matrix element $| \langle X^{1}\Sigma_{g}^{+}, v"=0 | D_{A-X} | 0_{u}^{+},v'=71 \rangle |^{2}$ is one of the largest one.

To summarize, this study shows that we can find an efficient transfer of the triplet molecules initially created by PA down to ground state molecules in their lowest vibrational level $v"=0$, relying on their excitation into a specific level of the $(2)^{3}\Pi_{g}$ state. Our model involves the $v'=12$ level, but this could slightly vary depending on the availability of new spectroscopic investigations, without changing the main conclusion of the proposal. Then the two-step spontaneous decay down to the $v"=0$ levels looks promisingly efficient, despite its apparent complexity induced by the strong SO coupling in the Cs$_2$ molecule.

\section{Discussion}
\label{sec:discussion}

\begin{figure}
    \begin{center}
    \includegraphics[width=1\linewidth]{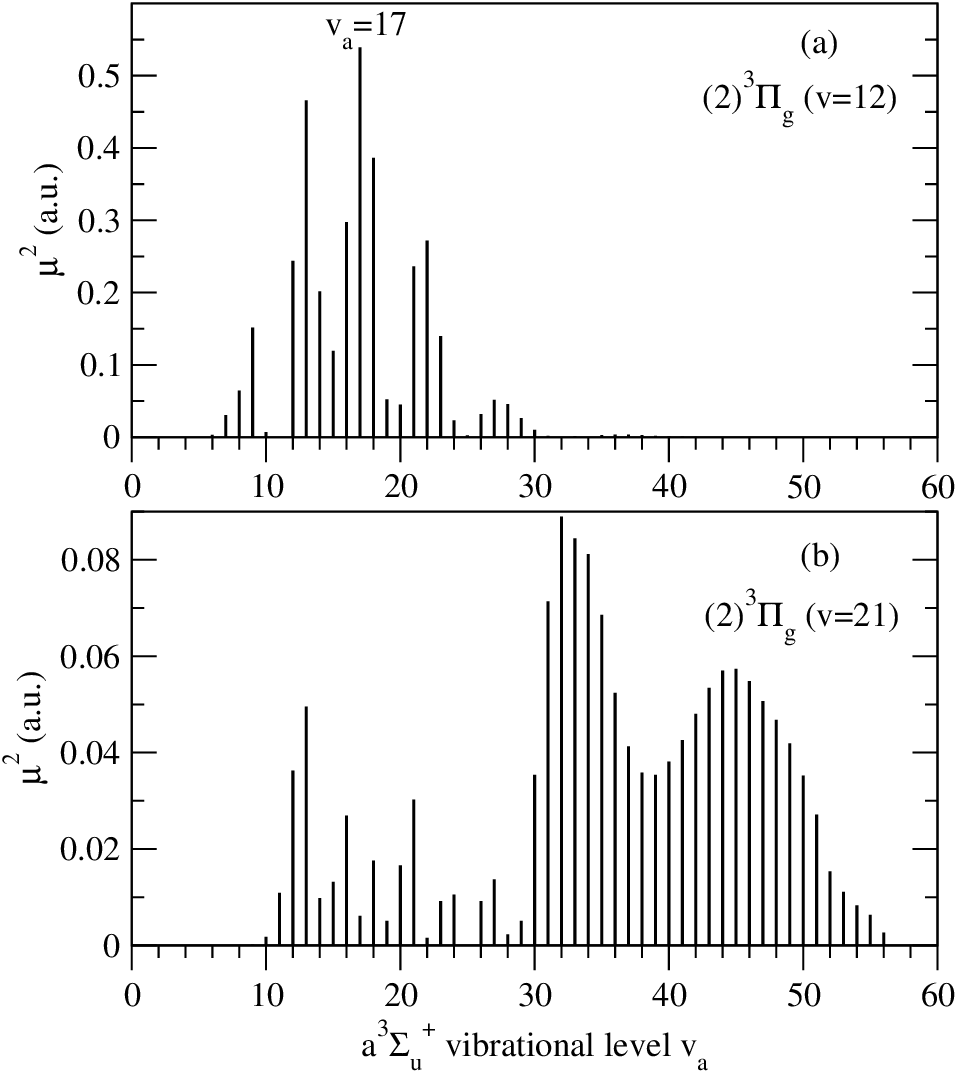}
    \end{center}
    \caption{Squared matrix element $\mu^2=|\langle (2)^{3}\Pi_{g}, v|D_{(2)-a}|a ^{3}\Sigma_{u}^{+}, v_a \rangle|^2$ (in atomic units) of the transition dipole moment function $D_{a-(2)}$ (see Fig.\ref{fig:transfer_scheme}d). (a) $v=12$; (b) $v=21$.}
     \label{fig:tPiv6_to_a}
\end{figure}

As shown in Fig \ref{fig:transfer_scheme}, the single-step decay of the $(2)^{3}\Pi_{g}$ levels back to $a ^{3}\Sigma_{u}^{+}$ levels obviously competes with the above conversion process. The rate depending on the squared matrix element $| \langle a ^{3}\Sigma_{u}^{+},v_a | D_{(2)-a} | (2)^{3}\Pi_{g},v \rangle |^{2}$ is evaluated according to Eq. \ref{eq:rate}, and is represented in Fig.\ref{fig:2tPig_relax}b. This rate is only twice as large as the $(2)^{3}\Pi_{g}$-$0_{u}^{+}$ rate, which confirms that both processes will indeed take place. The $v=12$ level (Fig.\ref{fig:tPiv6_to_a}a) then decays significantly back to the $v_a=17$ level, which could probably be partly repumped during the conversion process, depending on the details of the experimental procedure. Also, we have not considered in our model the $1_u$ and $2_u$ components of the $b$ fine structure manifold for the spontaneous decay of the $(2)^3\Pi_g$ state used for the conversion. These states represent a statistical weight of 80\% within this manifold, which may reduce the computed conversion efficiency. On the other hand, the only possible channel for these states to decay is the $X$ state, either through their admixture of $(1)^1\Pi_u$ state (for the $1_u$ component), or through weaker couplings (for the $2_u$ state).

This conversion process is in principle valid even if we start from high-lying vibrational levels of the $a$ state, obtained for instance when the PA laser is not exciting the specific giant resonances of the $0_g^-(6^2S+6^2P_{3/2})$ state of Cs$_2$, or when molecules are formed by MA. This is illustrated in Fig.\ref{fig:a_to_Pi}b where a quite low level ($v=21$) of the $(2)^{3}\Pi_{g}$ can be populated from the highest vibrational  $v_a=54$ level in the $a$ state. However the transition probability (proportional to the squared matrix element represented in Fig.\ref{fig:a_to_Pi}) is about 100 times smaller than for reaching the $v=12$ level from low lying $a$ levels, due to the large radial extension of the $v_a=54$ wave function with a weak amplitude in the short range. The other steps of the process are comparable to the previous case. Figure \ref{fig:Piv6_to_0+system}b shows the distributions of the populated $0_u^+$ levels, which again reflects the oscillatory structure  of the upper $v=21$ function. The $v'=71$ level is about twice less populated than from the $v=12$ level. As expected finally, the one-step decay of the $v=21$ level (Fig. \ref{fig:tPiv6_to_a}) populates individual $a$ levels with a probability about 5 to 10 times smaller than the $v=12$ decay , but generates a broader distribution of $a$ levels.

To minimize the effect of relaxation back to the ground triplet $a$ state, molecules can be optically pumped into the $0_{u}^{+}$ system by applying additional laser excitation to the $a^{3}\Sigma_{u}^{+}$ state. A setup employing a broadband laser light (as in ref. \cite{Sofikitis2009A}) can cover $v_{a}=15-25$ levels with the most of the molecule population. This process would transfer more molecules into the $0_{u}^{+}$ system effectively increasing the triplet-singlet conversion and accumulation of molecules in the ground $X$ state.

Additionally, the proposed triplet-singlet conversion probably explains the suppression of vibrational cooling of the $a$ state \cite{Sofikitis2010}.  As the $(2)^{3}\Pi_{g}$ state was used for the cooling transition, our work shows that a large fraction of the excited molecules are most likely lost at every absorption-emission cycle of the cooling process due to the two-step decay to the $X$ state.

The proposed conversion scheme is currently implemented experimentally at Laboratoire Aim\'e Cotton. The transitions involved in this conversion are within an easy reach for laser light. Moreover, the outlined process relies on spontaneous emission, but larger rates
could be obtained with coherent transfer  such as STIRAP, while molecules may be prepared in a well-defined internal ground state. This kind of conversion process would actually create a molecular cold gas suitable for the study of cold collisions between atoms and molecules, or among molecules, prepared in a well-defined internal state. As the number of cold molecules is quite large after the initial PA step \cite{Drag2000}, accumulating them in the $v=0$ level of the ground state would also provide an ideal starting point for the study of cold atom-molecules photoassociation into cesium trimers.

\section*{Acknowledgements}
This work is supported by the "Institut Francilien de Recherche sur les Atomes Froids'' (IFRAF).
M.P. acknowledges sabbatical support at the Laboratoire Aim\'{e} Cotton from Goucher College,
and support from the RTRA network "Triangle de la Physique''.

%

\end{document}